\documentclass[sigconf]{acmart}
\pdfoutput=1
\usepackage{subfigure}
\usepackage{multirow}
\usepackage{tabularx}
\usepackage{mathrsfs}
\usepackage{amsmath}

\usepackage{background}
\usepackage{lipsum}
\SetBgContents{Accepted by DLP-KDD 2022, https://dlp-kdd2022.github.io/}
\SetBgScale{1}
\SetBgAngle{0}
\SetBgOpacity{1}
\SetBgColor{black}
\SetBgPosition{current page.south}
\SetBgHshift{3cm}
\SetBgVshift{1cm}

\AtBeginDocument{%
  \providecommand\BibTeX{{%
    \normalfont B\kern-0.5em{\scshape i\kern-0.25em b}\kern-0.8em\TeX}}}

\begin{document}

\title{A Brief History of Recommender Systems}

\author{ Zhenhua Dong$^1$, Zhe Wang$^{2,3}$, Jun Xu$^4$, Ruiming Tang$^1$, Jirong Wen$^4$}

\affiliation{\institution{
$^{1}$ Huawei Noah's Ark Lab; 
$^{2}$ ByteDance Ltd; 
$^{3}$ Tsinghua University\\
$^{4}$ Gaoling School of Artificial Intelligence, Renmin University of China \country{China}\\
\{dongzhenhua,tangruiming\}@huawei.com; 
zhe.wangz@bytedance.com;zhewang06@tsinghua.edu.cn; 
\{junxu,jrwen\}@ruc.edu.cn\\
}
}

\renewcommand{\shortauthors}{Zhenhua Dong, et al.}

\begin{abstract}
Soon after the invention of the World Wide Web, the recommender system emerged and related technologies have been extensively studied and applied by both academia and industry. Currently, recommender system has become one of the most successful web applications, serving billions of people in each day through recommending different kinds of contents, including news feeds, videos, e-commerce products, music, movies, books, games, friends, jobs etc. These successful stories have proved that recommender system can transfer big data to high values. This article briefly reviews the history of web recommender systems, mainly from two aspects: (1) recommendation models, (2) architectures of typical recommender systems. 
We hope the brief review can help us to know the dots about the progress of web recommender systems, and the dots will somehow connect in the future, which inspires us to build more advanced recommendation services for changing the world better.

\end{abstract}





\maketitle

\section{Introduction}
In 1992, Belkin and Croft ~\cite{belkin1992information} analyzed and compared information filtering and information retrieval, where the information retrieval is the fundamental technology of search engine, and the recommender system is mainly based on the technology of information filtering. In the same year,~\citet{goldberg1992using} proposed Tapestry system which is the first information filtering system based on collaborative filtering through human evaluation. Inspired by the study, some researchers from Massachusetts Institute of Technology (MIT) and University of Minnesota (UMN) developed the news recommendation service, named GroupLens~\cite{resnick1994grouplens}, whose key component is a user-user collaborative filtering model. Prof. John Riedl founded a research lab at UMN, also named GroupLens, which is the pioneer of recommender system studies. For music and video, similar recommendation technologies have been applied by the Ringo system~\cite{shardanand1995social} and Video Recommender~\cite{hill1995recommending}, respectively. Along with the emergence of e-commerce, the industry realized the business value of recommendation. Net Perceptions~\cite{NetPerception}, the first company that focuses on offering the marketing recommender engine, was found in 1996. The customers include Amazon, Best Buy, and JC Penney etc. ~\citet{schafer1999recommender} explained how recommender systems help E-commerce sites to increase sales through analyzing six web sites, from three aspects such as interfaces, recommendation model, and user inputs. Since then, the academic studies and industrial practical applications became the two wheels of the progress of recommender system technologies. In the fall of 1997, the GroupLens research lab launched the MovieLens ~\cite{harper2015the} project and trained the first version of the recommender model with the EachMovie dataset. After that, several MovieLens datasets were continuously released during 1998 to 2019, and became one of the most popular datasets for recommendation studies. 

From the perspective of recommendation models, collaborative filtering technologies dominated the recommender system applications and studies before 2005, such as user-user collaborative filtering ~\citeN{herlocker1999an,breese1998empirical}, item-item collaborative filtering ~\citeN{linden2003amazon,sarwar2001item} and Singular Value Decomposition (SVD) based collaborative filtering ~\cite{sarwar2000application}. Inspired by the Netflix Prize during 2006 to 2009, the matrix factorization models had been extensively studied~\citeN{koren2008factorization, koren2010collaborative}. In the same period, some researchers began to propose informal arguments that the evaluation of the recommender systems should move from the conventional accuracy metrics to the user-centric evaluation~\cite{mcnee2006being}.

Because of the fast development of basic research and commercial applications in recommender system, the community decided to hold the first ACM Recommender Systems Conference ~\cite{konstan2007proceedings} at UMN in 2007. Currently, ACM RecSys has become one of the most important annual academic conference that focuses on the study of recommender system. In the same year, \citet{richardson2007predicting} presented a logistic regression (LR) model that achieved a 30\% reduction in term of the error in click-through rate estimation. Since then, LR models were continuously improved from different aspects, including optimizing methods~\citeN{graepel2010web,mcmahan2013ad}, automatic feature engineering~\cite{he2014practical} etc. In 2010,~\citet{rendle2010factorization} proposed the Factorization Machines (FMs) which combines the advantages of Support Vector Machines (SVM) and factorization models. Based on FMs,~\citet{juan2016field} proposed Field-aware Factorization Machines (FFMs) which considers the fields of features when modeling the weight of each feature pair. In the meantime, more and more studies pay attention to the user experiences in recommender system.~\citet{pu2011a, pu2012evaluating} proposed a user-centric evaluation framework for recommender systems; ~\citet{konstan2012recommender} appeal the evolution of the recommender system study from research concentrated purely on algorithms to research concentrated on user experience.

Since 2016, recommendation models based on deep neural networks have emerged in both academia and industry. As for the industrial recommendation models, the Wide\&Deep model~\cite{cheng2016wide} and DeepFM~\cite{guo2018deepfm} had been deployed for improving the App recommendations. YouTubeDNN \cite{covington2016deep} and correct-sfx \cite{yi2019sampling} were used for increasing the accuracy of video recommendation. DIN \cite{zhou2018deep} and DIEN \cite{zhou2019deep} were proposed for modeling the sequential information like user interests with attention mechanism. Wang et al. proposed DCN \cite{wang2017deep} and DCN V2 \cite{wang2021dcn} to automatically and efficiently learn bounded-degree predictive feature interactions.
In academia, researchers also proposed important deep recommendation models, such as FNN \cite{zhang2016deep}, PNN \cite{qu2016product}, NeuralCF \cite{he2017neuralcf}, NFM \cite{he2017neural}, CVAE \cite{li2017collaborative}. To address the issue of reproducibility in recommendation model studies~\citeN{dacrema2019are, lin2019the}, ~\cite{zhu2020fuxictr} developed an open benchmark for CTR prediction, named FuxiCTR. ~\citet{sun2020are} created benchmarks, like reproducible and fair evaluation metrics, for implicit-feedback based top-N recommendation algorithms. ~\cite{zhao2020recbole} proposed a unified framework to develop and reproduce recommendation algorithms for the research purpose. There are also some other open source recommendation models which largely advanced the progress of recommender system studies~\citeN{guo2015librec, NeuRec, ekstrand2011lenskit}.

In recent years, for addressing the biases in recommender systems, there are increasingly studies on causal inference inspired recommendation~\citeN{yuan2019improving, dong2020counterfactual}. \citet{schnabel2016recommendations} provided an approach to handling the selection biases by adapting models and estimations from causal inference. Thorsten taught a course named counterfactual machine learning~\cite{CF18} in 2018. Most of the course content was based on examples from information retrieval and recommender systems.
In this paper, we do not attempt to give a comprehensive review of all aspects of recommender systems, such as human computer interaction \citeN{cosley2003seeing, swearingen2001beyond}, evaluation \citeN{beel2013research, del2008evaluation, pu2011a}, privacy \citeN{zhan2010privacy, aimeur2008lambic}, attacking \cite{lam2004shilling}, user experiences \citeN{konstan2012recommender, pu2011a}, fairness \cite{ekstrand2019fairness} etc. There have been a number of thorough surveys about these related research topics. In the rest of the studies, we mainly review two aspects of recommender systems based on our research and industrial experiences, including practical recommendation models and the architectures of typical recommender systems, corresponding to Section 2, 3. Finally, we will briefly discuss some ideas about the future recommender system.

\section{Recommendation Models}
There have been a number of surveys and books focusing on the recommendation models \citeN{ricci2011introduction, zhang2019deep, ekstrand2011collaborative}, in this session, we will introduce five kinds of practical recommendation models chronologically. The models are selected mainly based on two principles: the model has important effect on the progress of recommender system, and the model has been successfully deployed on the products. 

\begin{figure*}[!thbp]
	\centering
	\includegraphics[scale=0.25]{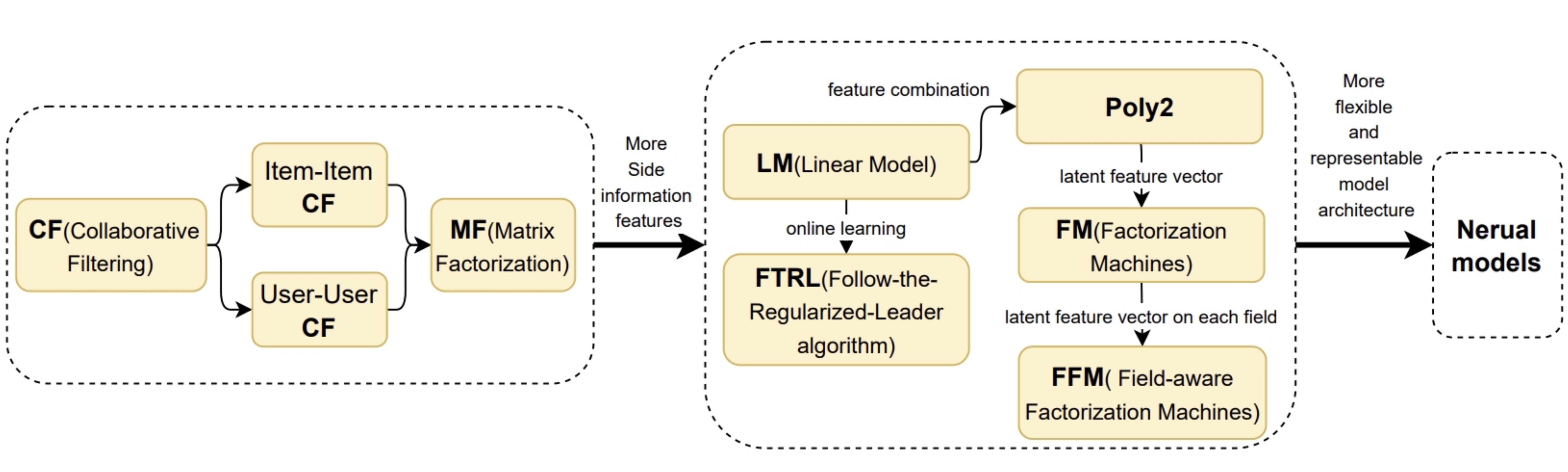}
	\caption{
	Evolution graph of classic recommendation models
	}
	\label{fig:EvoluationClassic}
\end{figure*}

\subsection{Collaborative Filtering}
Most of early recommendation models belong to collaborative filtering technologies, K-Nearest-Neighbor (KNN) models predict the user neighbors of a target user u through computing a similarity between u's prior preferences and the preferences of other users, then the preferences of the neighbors assigned to a target item i can be as a prediction for u to i \cite{herlocker1999an}, this classic recommendation model is known as user-user based collaborative filtering. With the similar way, item-item based collaborative filtering predicts a user's preference to an item i based on the user's preferences on the neighbors of i. When the number of users is more than that of items, item-item based collaborative filtering model is faster than the user-user one \cite{sarwar2001item}. Some studies \cite{wang2006unifying} combine the two kinds of collaborative filtering approaches. To improve the efficiency of recommendation based on big data, the commercial recommender systems applied SVD \cite{sarwar2002incremental} for reducing the dimensionality of user-item preference matrix to a latent taste dimensions.
During the Netflix Prize, the Matrix Factorization (MF) models \cite{koren2009matrix} had been deeply studied for rating prediction task, the MF is another kind of collaborative filtering technology, which is better than the classic KNN models by considering factors like implicit feedback and temporal information.
It is interesting to find that the Netflix product did not really use the winner's solution, which uses the ensemble approach for combining different models, too large engineering effort for implementation is one reason \cite{ricci2015recommender}.
Many recommender systems can only collect positive user feedbacks, so several one-class collaborative filtering models \citeN{pan2008one, sindhwani2010one} are proposed for such recommendation scenarios, where some missing samples as negative. Yu et. al. \cite{yu2017selection} develop efficient optimization techniques to model "full" samples, where all missing entries as negative samples.

Collaborative filtering based recommendation model is easy to implement and deploy, but there are also some problems, like cold start issues for both user and item, low computing efficiency when the numbers of users and items are huge. 

\subsection{Linear models}
Linear models can mitigate the cold start problem through considering the side information, such as user demographic, historic behaviors, item attributes, contextual information.
Logistic regression is one of the most practical linear models for recommendation, especially for the click-through rate (CTR) task \citeN{he2014practical}. Since it can naturally model the value of CTR between 0 and 1. Given a data set with \textit{$n$} instances (\textit{$y_i$, $x_i$}), i = 1,...,\textit{$m$}, where \textit{$y_i$} is the label and \textit{$x_i$} is the feature vector, and the model \textit{$w$} can be obtained by learning the following optimization: 

\begin{equation}\label{1.1}
\min_{\boldsymbol{w}} \frac{\lambda}{2} \|\boldsymbol{w}\|_2^2+\sum_{i=1}^{n}\log(1+\exp(-y_i\phi(\boldsymbol{w},\boldsymbol{x}_i))),
\end{equation}
where $\lambda$ is the regularization parameter and the $\|\boldsymbol{w}\|_2^2$ term is used to avoid overfitting data. The second term in $(\ref{1.1})$ is an approximate sum of training errors by using the logistic loss. The $\phi$ function is important. Traditionally, a linear model (LM) is considered with $\phi(\boldsymbol{w}, \boldsymbol{x})=\boldsymbol{w}^{T}\boldsymbol{x}$. Since every feature corresponds to an individual weight of $w$, the jointly weight between features is not taken into account. So a sort of feature engineering technologies are proposed for modeling the feature pair conjunction. Beside artificial feature interactions, degree-2 polynomial ($Poly2$) \cite{chang2010training} is a widely adopted model, which learns a weight for every possible feature pair. Decision tree plus LR is another practical automatic feature engineering approach, Facebook \cite{he2014practical} designs a hybrid model structure, where boosted decision tree transforms the input features, and the output of each individual tree can be as the input feature to a logistic linear classifier.
There are some important progresses about optimizing algorithms for LR, \cite{graepel2010web} learns the LR with Bayes optimization for improving the Ads CTR estimator of Microsoft Bing. McMahan et. al. \cite{mcmahan2011follow} propose the Follow-the-Regularized-Leader (FTRL) algorithm, and then implement an FTRL-Proximal online learning algorithm for improving the accuracy of CTR prediction task \cite{mcmahan2013ad}.

Logistic regression model is practical for industry recommender system, since it is good at modeling the side information and easy to extend. But LR can not model features conjunction by itself, the conjunction operation always depends on artificial feature engineering or other models, like $Poly2$ or boost decision trees, so features conjunction and model optimization are two isolate stages. 

\subsection{Low rank models}
Generalized low rank models \cite{udell2015generalized} can approximate a tabular data set by a low rank representation. 
Matrix factorization and Factorization Machines (FMs) are two popular low rank models, the former one had been introduced in the previous section, while the FMs \cite{rendle2010factorization} provides another feature conjunction method, which can decomposes one weight of $Poly2$ for each feature pair into the inner product of two $k$ dimensional latent vectors corresponding to each feature in the pair, and the $\phi$ function in $(\ref{1.1})$ is as following:
\begin{equation}
\label{fm}
\phi_{\text{FM}}(\boldsymbol{w},\boldsymbol{x}) = \sum_{j_1=1}^{n}\sum_{j_2=j_1+1}^{n} (\boldsymbol{w}_{j_1}\cdot\boldsymbol{w}_{j_2})x_{j_1}x_{j_2},
\end{equation}

where $j_1$ (or $j_2$) are feature indexes, $\boldsymbol{w}_{j_1}$ (or $\boldsymbol{w}_{j_2}$) $\in \mathbb{R}^k$. Because $\boldsymbol{w}_{j_1}$ and $\boldsymbol{w}_{j_2}$ are not solely decided by the co-occurrence of feature pair $x_{j_1}$ and $x_{j_2}$, so FMs can mitigate the issue of insufficiently learning of rare feature pairs. Field-aware Factorization Machines \cite{juan2016field} groups features to several fields, and each feature has different latent vector for each field, and the $\phi$ is as following:
\begin{equation}
\label{ffm}
\phi_{\text{FFM}}(\boldsymbol{w},\boldsymbol{x}) = \sum_{j_1=1}^{n}\sum_{j_2=j_1+1}^{n} (\boldsymbol{w}_{j_1,f_2}\cdot\boldsymbol{w}_{j_2,f_1})x_{j_1}x_{j_2},
\end{equation}

where $f_1$ and $f_2$ are the fields of $j_1$ and $j_2$ respectively. When deciding the weight for $x_{j_1}x_{j_2}$, we use ${j_1's}$ corresponding latent vector on ${j_2's}$ field and ${j_2's}$ corresponding latent vector on ${j_1's}$ field. FFMs  won some CTR prediction competitions, like Criteo and Avazu Kaggle competitions. \cite{chang2020autoconjunction} indicated that $Poly2$ tends to be good at modeling the dense features, while the FM/FFM are useful for sparse ones, so a model-based feature conjunction method named AutoConjunction has been proposed.

Fig.\ref{fig:EvoluationClassic} summarizes the evaluation of the above classic recommendation models. Feature conjunction is important to the accuracy of recommendation estimator. Although the above models can automatically learn the first and second order feature conjunctions efficiently, it is still hard to extend them to model the combined features more than two. So researchers began to study the neural recommendation models, whose structure are more flexible for advanced feature conjunction.

\subsection{Neural models}
\begin{figure*}[!thbp]
	\centering
	\includegraphics[scale=0.3]{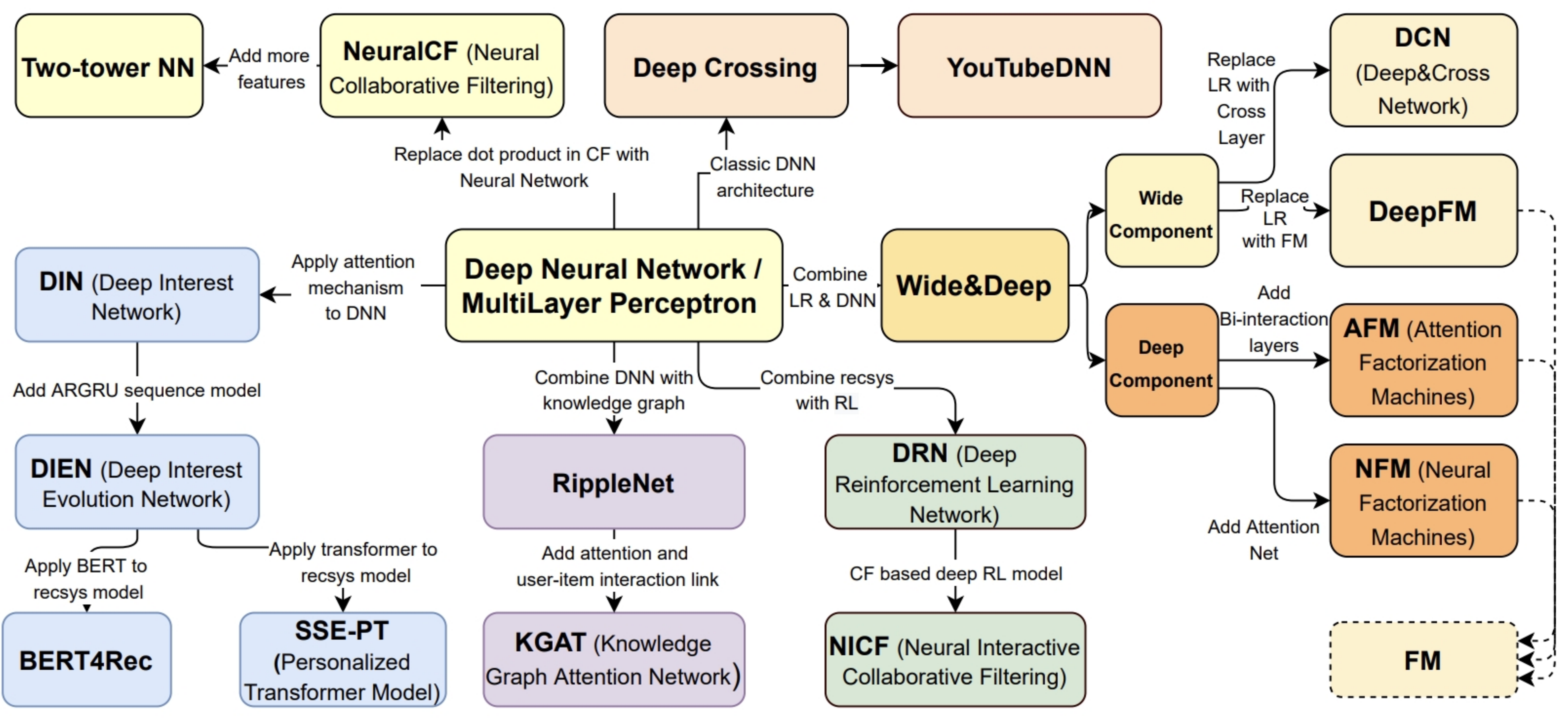}
	\caption{
	Evolution graph of deep learning recommendation models
	}
	\label{fig:EvolutionDNN}
\end{figure*}

Since 2016, deep neural network based recommendation model has received rapidly growing attention. It not only became research hotspot in academia, but also dominated development of industry recommender system and brought huge commercial value. Starting from general Multi-layer Perceptron, Fig.\ref{fig:EvolutionDNN} illustrates main evolution process of deep learning recommendation model which includes the following directions. 


\textit{Classic deep neural network}. Deep Crossing~\cite{shan2016deep} is an early-stage deep learning recommendation model, while its architecture is classic and has been inherited by many following models. Deep Crossing has three main components, such as embedding layer, stacking layer and residual layers. The function of embedding layer is to transform sparse categorical features into dense embeddings. Stacking layer is used for concatenating embedding and other continuous features. The last residual layers consisting of residual units automatically combine features to produce a superior model. YouTubeDNN~\cite{covington2016deep} is another influential model with similar architecture as Deep Crossing. The main difference is that YouTubeDNN uses ReLU neurons instead of residual units as the feature crossing layers. The most valuable thing of YouTubeDNN is that it discloses some details of industry recommendation model, such as feature processing, specific methods of model training and serving.

\textit{Two-tower neural network}. Since the nature of recommendation is to calculate a user interest to a specific item, two-tower neural network containing user tower and item tower becomes a straight forward solution for recommendation model. CF based matrix factorization is a classic and the simplest two-tower model using dot product to calculate similarity between item and user. NeuralCF~\cite{he2017neural} replaces dot product with MLP to enhance feature crossing capacity of the model. Some follow-up studies like YouTube Neural Retrieval Model~\cite{yi2019sampling} and Embedding-based Retrieval model of Facebook Search~\cite{huang2020embedding} add more features to each tower in the basic two-tower structure. The most important benefit of two-tower structure for industry application is highly efficient model serving. By making use of ANN (Approximate Nearest Neighbor) search in user and item embedding space, recommender system can fetch hundreds of relevant items from millions level candidates pool with a constant time complexity, which is necessary for industry online system.

\textit{Shallow and deep model}. This type of model mainly refers to the Wide\&Deep model~\cite{cheng2016wide} and its subsequent variants, such as Deep\&Cross~\cite{wang2017deep}, DeepFM~\cite{guo2018deepfm}, AFM~\cite{xiao2017attentional} etc. The main idea is to combine two or more deep learning networks with different advantages to enhance model capability. The original Wide\&Deep model is composed of LR and DNN, so it has both advantages of strong memorization of LR and generalization of DNN. To strengthen capability of feature crossing, DeepFM replaces LR with FM as a new wide component, while cross layer takes the place of LR in Deep\&Cross. On the other hand, AFM and NFM~\cite{he2017neural} improve deep component by adding attention net and bi-interaction layers.

\textit{Natural Language Processing (NLP) inspired neural model}. There are some NLP key technologies have been applied for improving the performance of recommender system. 
DIN~\cite{zhou2018deep} introduces attention mechanism into recommendation model to learn relevance between user historically interactive items and target item. DIEN~\cite{zhou2019deep} uses sequence model to learn the evolution trend of user interest. 
Transformer models, like BERT~\cite{devlin2018bert}, have achieved great successes for NLP tasks, they are also quickly applied in the field of sequential recommendation, like BERT4Rec~\cite{sun2019bert4rec} and SSE-PT~\cite{wu2020sse}.

\textit{Deep Reinforcement learning}. The environment of online recommender system is a dynamic feedback loop. The user’s interactive behaviors are collected in real time and used to the model training process. Reinforcement learning is good at modeling this kind of continuously changing environment.
Some early studies on Exploration and Exploitation (E\&E), like LinUCB~\cite{li2010contextual}, try to use real-time rewards to improve model. Some studies combine deep learning and reinforcement learning, and enhances the recommendation performance by letting the recommendation model learn user rewards in real time. Among these studies, DRN (Deep Reinforcement Learning Network)~\cite{zheng2018drn} is a deep Q-Learning based recommendation framework, whose main innovation is to explore model parameter online by using dueling bandit gradient descent. NICF (Neural Interactive Collaborative Filtering)~\cite{zou2020neural} is another typical 
model which can quickly catch the user's interests through representing the exploration policy with a two-tower neural network.

\textit{Graph neural network}. Much of internet data is in the form of graphs like knowledge graph, social network and user-item bipartite graph, so how to efficiently use graph structure data has become a key research direction for providing more accurate, diverse, and explainable recommendation.  Some random walk based graph embedding approaches, like DeepWalk~\cite{perozzi2014deepwalk} and Node2vec~\cite{grover2016node2vec} make initial trails on this direction. In recent years, different methods of graph neural network have been proposed to directly build a recommendation model on the graph structure. RippleNet~\cite{wang2018ripplenet} starts from selected seeding nodes and recursively propagates the embeddings from a node’s neighbors to refine the node’s embedding. Furthermore, KGAT (Knowledge Graph Attention Network)~\cite{wang2019kgat} adds user-item interaction links into graph and employs an attention mechanism to discriminate the importance of the neighbors based on RippleNet.

\subsection{Causality inspired methods}
Recently, the researchers in recommender system community have realized the importance of technology evolution from association learning to causal inference, which can not only improve the accuracy but also have potential to benefit the recommendation in a broader scope of perspectives, such as debias, transparence, fairness, evaluation, robustness, etc. Beyond modeling with the statically observed data, the causality based methods can actively explore new situations through intervention, or build and learn the unobserved world with counterfactual learning methods. For intervention technologies, \cite{bonner2018causal} makes use of uniform data to handle the bias problems through Inverse Propensity Score (IPS), Doubly Robust (DR), joint learning. \citeN{joachims2017unbiased, wang2018position} apply swap intervention approach to improve the ranking tasks. \citeN{fang2019intervention, agarwal2019estimating} propose several debiasing methods with natural intervention data, and prove their efficiency for mitigating the position bias of information system. For counterfactual learning, \cite{dong2020counterfactual} summarizes some counterfactual approaches for modeling three kinds of data, such as observed biased data, observed unbiased data and unobserved data. The counterfactual approaches mainly include counterfactual data learning, the correcting biased observed data and doubly robust methods. The basic idea of counterfactual data learning is that we counterfactually assume that each user expresses his or her preferences to each item, and construct the counterfactual samples. The full information data through combining the factual and counterfactual data has potential to learn a more unbiased model. \citeN{yuan2019improving, yuan2020unbiased, chen2020efficient} propose efficient methods to learn the parameters of recommendation model from the whole data (including all missing data). Yang et al. \cite{yang2021top} counterfactually simulate user ranking-based preferences to handle the data scarce problem with Pearl’s causal inference framework. IPS \cite{schnabel2016recommendations} is the most popular method of the correcting biased data approach, it is defined as the conditional probability of receiving the treatment given pre-treatment covariate \cite{rosenbaum1984reducing}, but IPS method should satisfy the prerequisite about overlap and unconfoundedness \cite{austin2011introduction}, which are not easy to achieve in recommendation scenario when candidates number is large. The doubly robust methods have IPS part and direct method part, corresponding to the first two counterfactual approaches. Dudik et al. \cite{dudik2011doubly} prove the DR can yields accurate prediction when either part is a consistent estimator for policy evaluation and learning.

Structural Causal Model (SCM) \cite{pearl2009causality} and Potential Outcome Model (POM) \cite{little2000causal} are two mainstream causal frameworks, most of the current causality inspired recommendation studies \citeN{zhang2021causal, liu2021mitigating, wang2021clicks} try to analyze the problems (e.g. biases) with SCM, and a few studies \cite{wang2020causal} begin to frame recommendation with POM for developing the deconfounded recommender.

\section{Typical Recommender Systems}
Beyond models, the practical recommender systems need other important components. In this section, we will introduce typical recommender systems and their architectures.

\textit{Collaborative Filtering based Recommender System.} Tapestry \cite{goldberg1992using} is the earliest collaborative filtering based recommender system, which relied on the opinions of people in a small community. The Fig2. in \cite{goldberg1992using} describes the flows of information through the major architectural components of Tapestry, including \textit{Indexer}, \textit{Document store}, \textit{Annotation store}, \textit{Filterer}, \textit{Little box}, \textit{Remailer}, \textit{Appraiser}, \textit{Reader/Browser}. The collaborative filtering of Tapestry can not server large communities, since we can not assume each user knows each others, so the \textit{neighbors} mechanism had been implemented by GroupLens \cite{resnick1994grouplens} news recommender system, whose architecture is illustrated by the Fig2. in \cite{resnick1994grouplens} and Fig2. in \cite{sarwar2000application}, where the \textit{Better Bit Bureau} is plugged entity to the netnews architecture, it can collect the user preferences from clients, predict the recommendation scores, and send the score to clients. The core components include \textit{WWW Server}, \textit{Dynamic HTML Generator} and \textit{Recommender System}.

\textit{Two-layer Recommender System.} 
Bambini et. al. \cite{ricci2011introduction} describes how to integrate a recommender system into a real production environment of one IP Television (IPTV) providers. The Fig 9.2 in \cite{ricci2011introduction} describes the architecture of the IPTV recommender system, and the Fig 9.3 illustrates the model training in \textit{batch} stage and prediction in \textit{real-time} stage. As far as we know, it is the first public recommender system whose architecture is two-stage architecture. The IPTV recommender system integrates both collaborative filtering models and content-based models for satisfying different kinds of recommendation scenarios. Hulu's technology blog \cite{HuluArch} proposes their recommender system architecture, which includes online and offline parts, the former one is in charge of generating recommendation results and contains five main modules, like \textit{User profile builder}, \textit{Recommendation Core}, \textit{Filtering}, \textit{Explanation} and \textit{Ranking}, as shown in Fig 1 in \cite{HuluArch}. While the online components rely on offline system,
whose components include \textit{Data Center}, \textit{Related Table Generator}, \textit{Topic Model}, \textit{Feedback Analyzer} and \textit{Report Generator}, as shown in Fig 2 in \cite{HuluArch}.

\begin{figure*}[!thbp]
	\centering
	\includegraphics[scale=0.25]{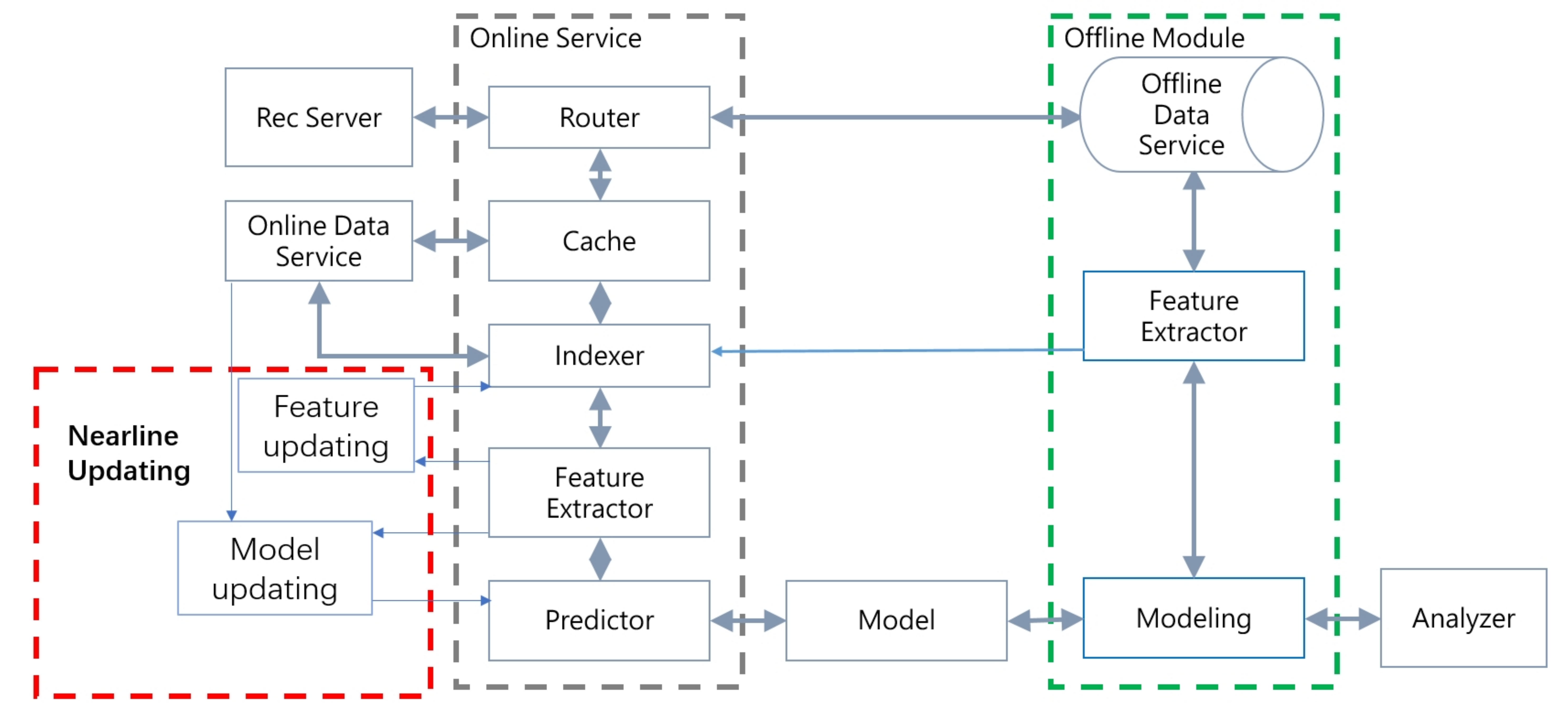}
	\caption{
	A general three-layer architecture diagram for industrial recommender systems. 
	}
	\label{fig:framework}
\end{figure*}

\textit{Three-layer Recommender System.}
Amatriain and Basilico \cite{ricci2015recommender} describe a generic three-layer architecture of recommender system at Netflix, where the \textit{Online Computation} can respond better to recent events, user interaction, and requests in real-time, like generating the recommendation results.
The \textit{Offline Computation} has more time to execute more complex computation with more data, like model training. While the \textit{Nearline Computation} is an intermediate between these two computations which can perform online-like computations but not in real-time, like user features updating. The main components of the three-layer architecture have been shown in Fig.11.10 in \cite{ricci2015recommender}. Dong ~\cite{MDM19} introduces a general three-layer architecture diagram for industrial recommender systems, and describes its key components. As shown in Fig.\ref{fig:framework}, there are mainly three layers, named \textit{Offline Module}, \textit{Online Service} and \textit{Nearline Updating}. The \textit{Offline Module} is in charge of model training. It extracts features from \textit{Offline Data Service}, then generate the samples through \textit{Feature Extractor}, and the recommendation model can be learned through the \textit{Modeling} component with batch learning mechanism. After the offline evaluation by \textit{Analyzer}, we select the best model and upload it to the \textit{Predictor} in \textit{Online Service}. The extracted features are also synchronized to the \textit{Indexer} of \textit{Online Service}. The \textit{Online Service} receives and distributes the user’s recommendation requirement through \textit{Router}. After parsing the requirement and acquire the user ID, we can get the feature vector x through the \textit{Indexer} and \textit{Feature Extractor}, finally, the \textit{Predictor} computes the recommendation score and generate the recommendation list. The \textit{Nearline Updating} is in charge of updating the features in real-time by \textit{Feature updating}, and updating the recommendation model according to specific scenario through \textit{Model updating}.

\section{About Future}
The future is very hard to predict, especially for the fast developing recommender systems and their widely applications. However, we hope to summarize some thoughts about the trends of recommendation based on the above history of web recommendation and our practical experiences. We think that the evolution of recommender systems mainly come from two driving forces, technological innovations and valuable industrial applications.

From the perspective of technological paradigm, the large pre-trained model is learned with big data based on self-supervised learning methods. It has potential to make use more past information to learn more efficient recommendation model. On the one hand, the recommender system can make use of the knowledge in the pre-trained model to improve the system efficiency; on the other hand, we also need to explore novel pre-training methods for recommendation tasks through improving the structure and objectives of the classic pre-train models. Human understands and interacts with the world through multimodal information, recommender system also needs to make use of the multimode technologies to model user, item and context. There are some studies \cite{lin2021m6, rahman2020integrating} about how to learn multimodal pre-train model.
Most of the previous recommendation work studied how to learn the associations with delicately complicated models through fitting the observed data. 
Although the association based learning methods achieve many successes, the disadvantages gradually emerge, such as the performance is unstable in out-of-distribution conditions, the complex models are nontransparent, the prediction result is less responsible. We think the neural science, causal inference and knowledge-enhanced methods have potential to inspire new solutions to handle the problems.

From the perspective of applications, recommender systems serve users, so more intelligent and user-centric applications will emerge, like naturally conversational recommender system and general intelligent assistants, would be the next mainstream recommendation applications. 
On the one hand, the future recommender system likes the British butler, who is well trained and serves people based on his knowledge and long serving experiences with the master; on the other hand, it also likes a knowledgeable teacher, who can help us to discover new useful information and benefit our lifelong growth. In the mean time, the user-centric and responsible evaluations would be future studied, we hope the future recommender system can provide more accurate, fair, accountable, knowledgeable, transparent and useful services for improving the user experiences in both long and short term.



\begin{acks}
Special thanks to Prof. John Riedl for his great and pioneering research in area of recommender systems, and his wonderful conversations with Zhenhua Dong about GroupLens, MovieLens, Net Perception during 2010 to 2011, there is no this work without him.
Thanks to the colleagues of Huawei Noah's Ark Lab, there is no this work without their fabulous work in both recommender system applications and academic studies in the past 10 years.
\end{acks}


\newpage
\bibliographystyle{ACM-Reference-Format}
\balance
\bibliography{ldc}

\end{document}